\documentclass[twocolumn,pra,twocolumn,superscriptaddress]{revtex4}
\usepackage{color}
\usepackage{amsmath}
\usepackage{amssymb}
\usepackage{graphicx}
\usepackage{tikz}

\usepackage{float}
\usepackage{bbm}

\usepackage{epsfig} 
\usepackage{verbatim}
\usepackage{array}
\usepackage{setspace}

\usepackage[unicode=true,
 bookmarks=true,bookmarksnumbered=false,bookmarksopen=false,
 breaklinks=false,pdfborder={0 0 1},backref=false,colorlinks=true]
 {hyperref}
\hypersetup{
 linkcolor=magenta, urlcolor=blue, citecolor=blue, pdfstartview={FitH}, hyperfootnotes=false, unicode=true}

\makeatletter
\@ifundefined{textcolor}{}
{%
 \definecolor{BLACK}{gray}{0}
 \definecolor{WHITE}{gray}{1}
 \definecolor{RED}{rgb}{1,0,0}
 \definecolor{GREEN}{rgb}{0,1,0}
 \definecolor{BLUE}{rgb}{0,0,1}
 \definecolor{CYAN}{cmyk}{1,0,0,0}
 \definecolor{MAGENTA}{cmyk}{0,1,0,0}
 \definecolor{YELLOW}{cmyk}{0,0,1,0}
}

\usepackage{amsfonts}\usepackage{tabularx}\usepackage{dcolumn}\usepackage{bm}\usepackage{graphicx}\usepackage{epstopdf}

\hypersetup{urlcolor=blue}

\makeatother
\newcommand\blue[1]{{\color{black}#1}}
\newcolumntype{C}[1]{>{\centering\arraybackslash$}p{#1}<{$}}

\begin{document}

\title{Tunable charge qubit based on barrier-controlled triple quantum dots}

\author{Xu-Chen Yang}
\affiliation{Department of Physics, City University of Hong Kong, Tat Chee Avenue, Kowloon, Hong Kong SAR, China, and City University of Hong Kong Shenzhen Research Institute, Shenzhen, Guangdong 518057, China}
\affiliation{Department of Physics, The University of Hong Kong, Pokfulam, Hong Kong SAR, China}
\author{Guo Xuan Chan}
\affiliation{Department of Physics, City University of Hong Kong, Tat Chee Avenue, Kowloon, Hong Kong SAR, China, and City University of Hong Kong Shenzhen Research Institute, Shenzhen, Guangdong 518057, China}
\author{Xin Wang}
\email{x.wang@cityu.edu.hk}
\affiliation{Department of Physics, City University of Hong Kong, Tat Chee Avenue, Kowloon, Hong Kong SAR, China, and City University of Hong Kong Shenzhen Research Institute, Shenzhen, Guangdong 518057, China}
\date{\today}

\begin{abstract}
We present a theoretical proposal of a tunable charge qubit, hosted in triple quantum dots. The manipulation is solely performed by changing the heights of the two potential barriers between the three dots, while the energy of all three dots are fixed. We have found that when the relative height of the two barriers are changed, the direction of the axis of rotation in performing single-qubit gates can be varied. On the other hand, the corresponding rotation speed can be tuned by raising or lowering the two barriers at the same time. Our proposal therefore allows for tunability of both the rotation axis and rotating speed for a charge qubit via all-electrical control, which may facilitate realization of quantum algorithms in these devices.
\end{abstract}

\maketitle

\section{introduction}

Electrons confined in semiconductor quantum dots are promising candidates for the physical realization of quantum computing. Either the charge \cite{Hayashi.03,Mizuta.17} or spin \cite{Loss.98,Levy.02,Petta.05,DiVincenzo.00} states of electrons can be used to encode a qubit. Perhaps the most intuitive realization is the double-quantum-dot charge qubit, for which an electron is allowed to occupy either one dot or the other, serving as the two logical states. Universal single-qubit operation can be performed by alternating between zero and large detuning (``tilt control'') \cite{Hayashi.03,Fujisawa.04,Fujisawa.06,Dovzhenko.11,Cao.13, Cao.15,Wang.17CQ}, which achieves $x$ and $z$-axis rotations, respectively. While this charge qubit has been demonstrated early-on to have very fast gate operations \cite{Hayashi.03,Shi.13}, it at the same time strongly suffers from charge noises \cite{Gorman.05,Petersson.10, Wang.17CQ},  which has limited its development. Spin qubits, on the other hand, are much less sensitive to charge noises \cite{Dial.13}. The last decade has witnessed accomplishments of very high control fidelities and long coherence times in single-qubit operations based on various systems \cite{Bluhm.10b,Bluhm.10c,Tyryshkin.11,Maune.12,Medford.13b,Pla.12,Pla.13,Muhonen.14,Veldhorst.14,Kawakami.16}. Nevertheless, spin qubits also come with a great deficiency that coherent two-qubit operations are challenging since couplings between two spin qubits \cite{Stepanenko.07,vanWeperen.11,Shulman.12,Trifunovic.12,Mehl.14b,Nichol.16} are typically weak.  This and other considerations have called a revival of interests on charge qubits, with the expectation that the strong Coulomb interaction between them may be suitable for realizing fast two-qubit gates while the charge noises are reduced by improvements in experimental techniques \cite{Shinkai.09,Li.15,Ward.16,Serina.17}. A successful example is the microwave-driven double-dot charge qubit \cite{Kim.15}, for which the qubit is operated essentially at zero detuning (a ``sweet-spot'' at which the exchange interaction is first-order insensitive to charge noises and causes a constant $z$-rotation \cite{Shim.16,Penfold.17}), and the relative phases of consecutive microwave bursts provide rotations around an arbitrary axis in the $xy$ plane (``microwave control''). Great tunability has also been recently demonstrated on certain types of ``hybrid qubits'' which are expected to combine advantages from manipulating both charge and spin states \cite{Shi.12,Kim.14,Wong.16,Thorgrimsson.17,Yang.17Wisc,Russ.16,Russ.17,Wang.16HYB,Cao.16,Wang.17HYB}.

The microwave control provides flexibility of choosing axes of rotation for a single qubit in the rotating frame, which greatly simplifies the control: it has been shown that if a sequence of piecewise constant pulses are to be used to implement the control on adjustable axes, only two pieces are sufficient to achieve arbitrary single-qubit rotation \cite{Shim.13}. Moreover, not only the rotation axes are flexible, the strength of the control field, i.e.~the rotating speed, can also be modulated by the amplitude of the microwave.  Together these advantages make the microwave control among the most viable control methods at present. Nevertheless, quantum gates based on microwave control are typically slow because the amplitude of the control fields are smaller as compared to energy level splitting in qubit devices \cite{Kim.15}. Application of microwave or similar radio-frequency fields can heat up the sample and can be hard to localize to a given qubit in a scaled-up array.
All-electrostatic operations are therefore still desired, because current technologies have achieved very high precision and short switching time in generating electrostatic control pulses, and that these pulse sequences are efficient in performing qubit control \cite{Petersson.10,Li.15}. These considerations have motivated us to find a qubit encoding scheme which can be controlled by all-electrical means that, at the same time, is flexible on both the direction of rotation axes and the rotating speed.

While the tilt control by changing the detuning is the most common ``electrical'' method to control a qubit, it has been recently realized, in double-quantum-dot spin qubit systems, that varying the barrier between the two dots (``barrier control'') \cite{Reed.16, Martins.16, Malinowski.17} serves as a powerful alternative to other methods, having advantages in many ways \cite{Yang.17b,Zhang.17,Yang.18,Shim.18}. In this paper, we apply the barrier control to a charge qubit encoded in a triple-quantum-dot system. We have found that, when the charge qubit is encoded using $(0,2,1)$ and $(1,2,0)$ states (entries in the brackets refer to electron occupancy in the respective dots), controlling the relative heights of the two potential barriers between the three dots provides ability to rotate the qubit around a wide range of axes on the $xz$ plane. At the same time, altering the potential barriers simultaneously with their relative height fixed changes the rotating speed. Our proposal therefore provides an example that flexible rotation axes and speed can be achieved with all-electrical control, which can be potentially useful in developing and controlling scalable quantum-dot qubit arrays.

The remainder of the paper is organized as follows. In Sec.~\ref{sec:model} we present the model, and in Sec.~\ref{sec:res} we show our results. We then conclude in Sec.~\ref{sec:conclusion}.

\begin{figure}[t]
(a)\centering\includegraphics[width=0.6\columnwidth]{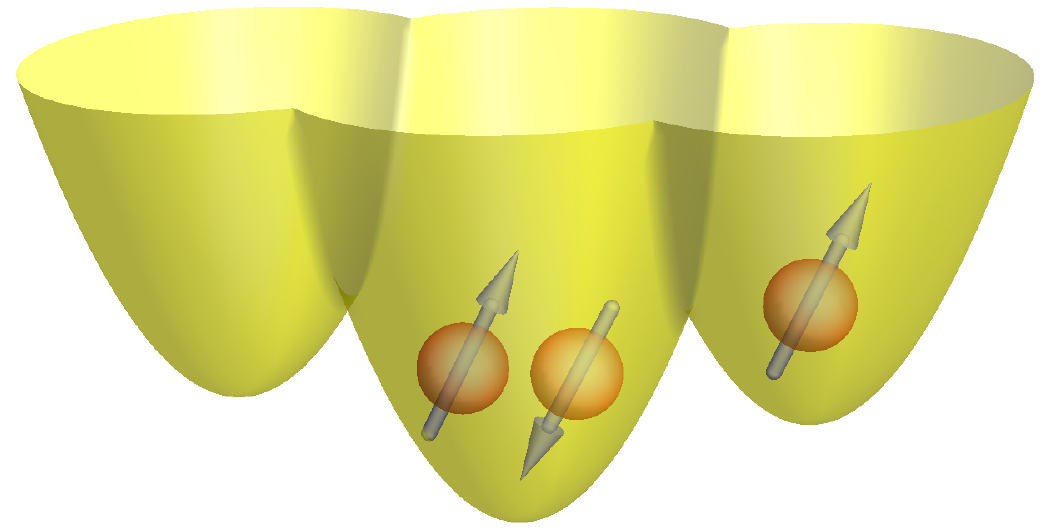}
\hspace{\fill}
(b)\centering\includegraphics[width=0.6\columnwidth]{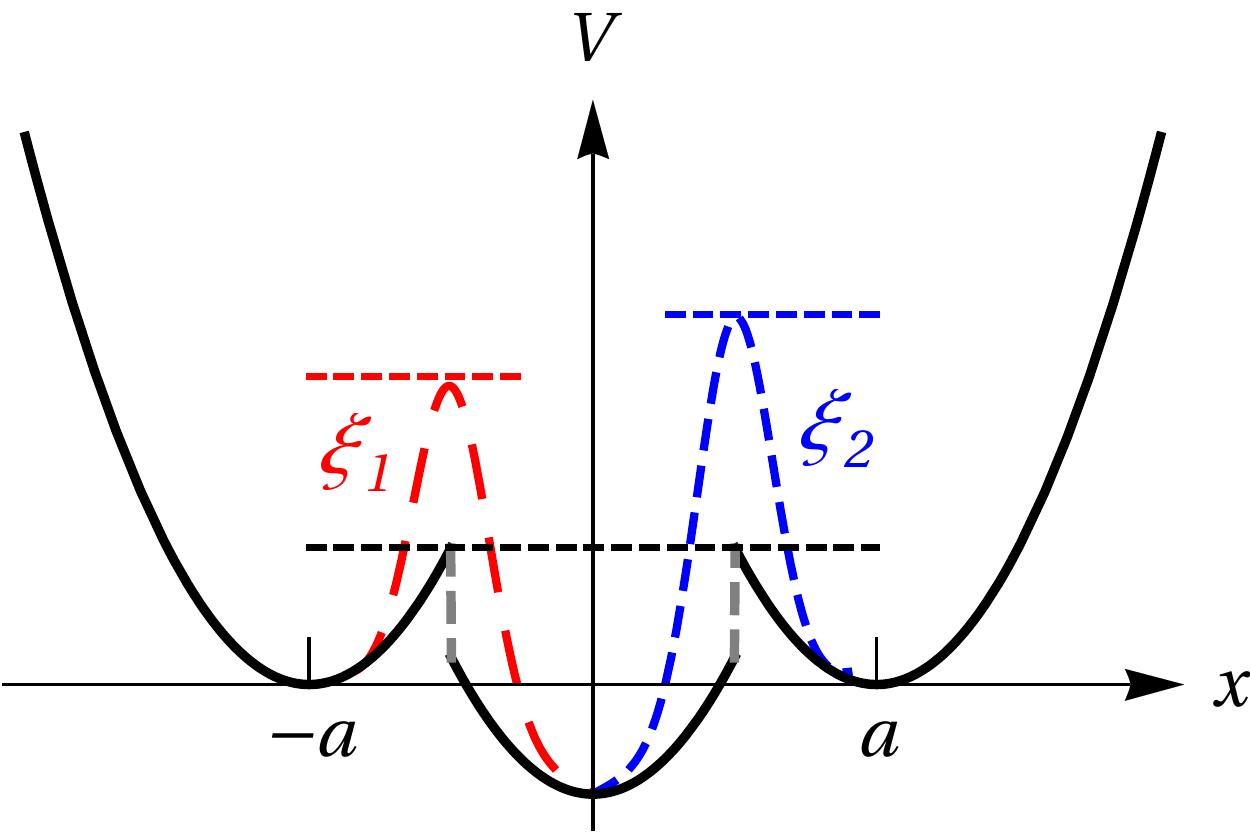}
\caption{(a) Schematic figure showing the triple quantum dots locating at $(x,y)=(\pm a,0)$ and $(0, 0)$. The middle dot is occupied by two electrons, while the third electron occupies either the left or the right dot.
(b) Schematic triple-well confinement potential under barrier control. The barrier control method changes the height of the potential barrier between the left and middle dots ($\Delta\xi_1$) and the one between the middle and right dots ($\Delta\xi_2$). }
\label{fig:tqd}
\end{figure}

\section{Model}
\label{sec:model}

We start with a lateral triple-quantum-dot system in the $xy$ plane, the Hamiltonian of which can be written as
\begin{equation}
H=H_s+H_I.
\label{eq:fullH}
\end{equation}
Here, the single-electron Hamiltonian $H_s=h(\bm{r}_1)+h(\bm{r}_2)+h(\bm{r}_3)$,
\begin{equation}
h(\bm{r})=\frac{1}{2m^*}\left[\bm{p}-e\bm{A}(\bm{r})\right]^2+V(\bm{r}),
\label{eq:sH}
\end{equation}
 where the effective electron mass $m^*=0.067m_e$, and the vector potential $\bm{A}$ implies a magnetic field along  $\hat{z}$.
$H_I$ includes the Coulomb interaction between three electrons in the system,
\begin{equation}
H_I=\sum_{1\le i<j\le3}\frac{e^2}{4\pi\kappa|\bm{r}_i-\bm{r}_j|}.
\label{eq:Hc}
\end{equation}
The confinement potential is defined by a sum of three parts,
\begin{equation}
V(x,y)=V_0(x,y)+G_1(x,y,\xi_1)+G_2(x,y,\xi_2).
\label{eq:V}
\end{equation}
The first part has a usual quadratic form for each dot,
\begin{eqnarray}
V_0(x,y)=
\begin{cases}
v_1, &x\le-a/2,\cr
v_2, &-a/2<x\le a/2,\cr
v_3, &x>a/2,
\end{cases}
\label{eq:threepiece}
\end{eqnarray}
where
\begin{equation}
v_i\equiv\frac{m^*\omega_0^2}{2}\left|\bm{r}-\bm{R}_i\right|^2+\mu_i
\label{eq:Vi}
\end{equation}
indicates the confinement potential of the $i$th quantum dot centering at $R_i$. The coordinates of $R_i$ are
\begin{equation}
\bm{R}_1=(-a, 0),\quad\bm{R}_2=(0, 0),\quad\bm{R}_3=(a, 0).
\label{eq:Vi}
\end{equation}
The remaining two terms of Eq.~\eqref{eq:V} are our control over the barriers between adjacent dots:
 \begin{equation}
 \begin{aligned}
G_1(x,y,\xi_1)&=\xi_1\exp\left\{-\frac{32\left[(x+a/2)^2+y^2\right]}{a^2}\right\},\\
G_2(x,y,\xi_2)&=\xi_2\exp\left\{-\frac{32\left[(x-a/2)^2+y^2\right]}{a^2}\right\}.
\end{aligned}
\label{eq:G}
\end{equation}
The height of the two barriers are controlled by parameters $\xi_1$ and $\xi_2$. We note that the two barriers are fixed at $(\pm a/2,0)$ and for this purpose, the quadratic part of the confinement potential, Eq.~\eqref{eq:threepiece}, has discontinuities at $x=\pm a/2$. The barrier functions $G_1$ and $G_2$  override the cusps between two adjacent quadratic potentials, and we use this setup because while we change the height of barriers we would like to minimize the other effects on the quantum-dot confinement potentials. It is also conceivable that experimentally one may change the barriers while leaving other factors characterizing the confinement potential, i.e. the location and energy of quantum dots, unchanged.  Schematic diagrams showing the confinement potential and the barrier control are presented in Fig.~\ref{fig:tqd}. In Fig.~\ref{fig:tqd}(b) we have used $\Delta\xi_1$ and $\Delta\xi_2$ to denote the change in the barrier heights.

In this work we apply the Hund-Mulliken approximation to solve the problem. We approximate the ground states by those of a harmonic oscillator:
\begin{equation}
\begin{aligned}
\phi_i(\bm{r})=\frac{1}{a_B^{}\sqrt{\pi}}\exp\left[{-\frac{1}{2a_B^2}\left|\bm{r}-\bm{R}_i\right|^2}\right],
\end{aligned}
\label{eq:ground}
\end{equation}
where the $a_B$ is the Fock-Darwin radius $\sqrt{\hbar/(m^*\omega_0)}$, and $i=1,2,3$ indicate the three quantum dots. We then orthogonalize the Fock-Darwin states in Eq.~\eqref{eq:ground} to obtain approximated single-electron wave functions in the triple-quantum-dot system. The orthogonalization is performed  by the transformation \cite{annavarapu2013singular,Yang.17}
\begin{equation}
\begin{aligned}
\left\{\psi_1,\text{ }\psi_2,\text{ }\psi_3\right\}^\text{T}=\mathcal{O}^{-1/2}\left\{\phi_1,\text{ }\phi_2,\text{ }\phi_3\right\}^\text{T},
\end{aligned}
\label{eq:trans}
\end{equation}
where $\mathcal{O}$ is the overlap matrix (defined as $\mathcal{O}_{l,l^\prime}\equiv\langle\phi_l|\phi_{l^\prime}\rangle$).

We consider three electrons (two spin-up and one spin-down) occupying the three quantum dots, and each dot allows a maximum of two electrons (i.e., only the lowest energy level is retained  in the Hund-Mulliken approximation). \blue{We note that keeping higher levels will not qualitatively change the results as the relevant states are far away from the computational subspace in energy. For a discussion on the effect of keeping three levels in each dot, see Appendix~\ref{appx:threelevels}.}  Our complete bases contain the following 9 states:

\begin{subequations}
		\begin{align}
	\left | \uparrow,\uparrow,\downarrow \right \rangle&=c^\dagger_{1\uparrow}c^\dagger_{2\uparrow}c^\dagger_{3\downarrow}|\text{vac}\rangle,\label{eq:9states1}\\
	\left | \uparrow,\downarrow,\uparrow \right \rangle&=c^\dagger_{1\uparrow}c^\dagger_{2\downarrow}c^\dagger_{3\uparrow}|\text{vac}\rangle,\\
	\left | \downarrow,\uparrow,\uparrow \right \rangle&=c^\dagger_{1\downarrow}c^\dagger_{2\uparrow}c^\dagger_{3\uparrow}|\text{vac}\rangle,\\	
	\left | \uparrow,\uparrow\downarrow,0 \right \rangle&= c^\dagger_{1\uparrow}c^\dagger_{2\uparrow}c^\dagger_{2\downarrow}|\text{vac}\rangle,\\
	\left | \uparrow\downarrow,\uparrow,0 \right \rangle&= c^\dagger_{1\uparrow}c^\dagger_{1\downarrow}c^\dagger_{2\uparrow}|\text{vac}\rangle,\\
	\left |0, \uparrow,\uparrow\downarrow \right \rangle&= c^\dagger_{2\uparrow}c^\dagger_{3\uparrow}c^\dagger_{3\downarrow}|\text{vac}\rangle,\\
	\left |0, \uparrow\downarrow,\uparrow \right \rangle&= c^\dagger_{2\uparrow}c^\dagger_{2\downarrow}c^\dagger_{3\uparrow}|\text{vac}\rangle,\\
	\left | \uparrow,0,\uparrow\downarrow \right \rangle&= c^\dagger_{1\uparrow}c^\dagger_{3\uparrow}c^\dagger_{3\downarrow}|\text{vac}\rangle,\\
	\left | \uparrow\downarrow,0,\uparrow \right \rangle&= c^\dagger_{1\uparrow}c^\dagger_{1\downarrow}c^\dagger_{3\uparrow}|\text{vac}\rangle\label{eq:9states9},
		\end{align}		
\end{subequations}
where $|\text{vac}\rangle$ refers to a vacuum state and $c_{i\sigma}^\dagger$ creates an electron on the $i$th dot with spin $\sigma$. Under these bases, the Hamiltonian can be expressed as a $9\times 9$ matrix,  {the elements of which are then obtained from the configuration interaction calculation \cite{fixedenergylevels}}. Diagonalization of the matrix gives the energy spectra and eigenstates of the system.

\section{Results}
\label{sec:res}

In this work we take $\mu_1=\mu_3=0$ while $\mu_2<0$, so the middle dot has lower energy than the other two.  {The physical dimensions of quantum dots are chosen to be $\hbar \omega_0=120$ $\mu$eV and $a=150$nm, which conform to the usual parameters used in experiments and theoretical calculation \cite{burkard1999coupled,li2010exchange,kouwenhoven1997excitation,zhang2018leakage}; larger $\hbar \omega_0$ is also possible, but since the same qualitative result is obtained, we only discuss in regards of $\hbar \omega_0 = 120$ $\mu$eV.} Two out of the nine states Eqs.~\eqref{eq:9states1}-\eqref{eq:9states9} stand out as possible ground states, which we label as our qubit states as
\begin{equation}
\begin{aligned}
\left|0\right \rangle &= \left | \uparrow,\uparrow\downarrow,0 \right \rangle,\\
\left|1\right \rangle &= \left | 0,\uparrow\downarrow,\uparrow \right \rangle.
\end{aligned}
\end{equation}
This is essentially a charge qubit.
In our calculation we have also found that hybridization between the qubit states and other states are very small, so we will only be concerned with the qubit states while discussing the results. In this work, we consider a situation where $\mu_1$, $\mu_2$, and $\mu_3$ are fixed so our control is exercised solely via the two barriers, $\xi_1$ and $\xi_2$. We shall show that the control of the two barriers allows for flexibilities in both the rotation axis and the rotating speed, which adds tunability to the charge qubit traditionally conceived.

\begin{figure}[tbp]
(a)\centering\includegraphics[width=0.7\columnwidth]{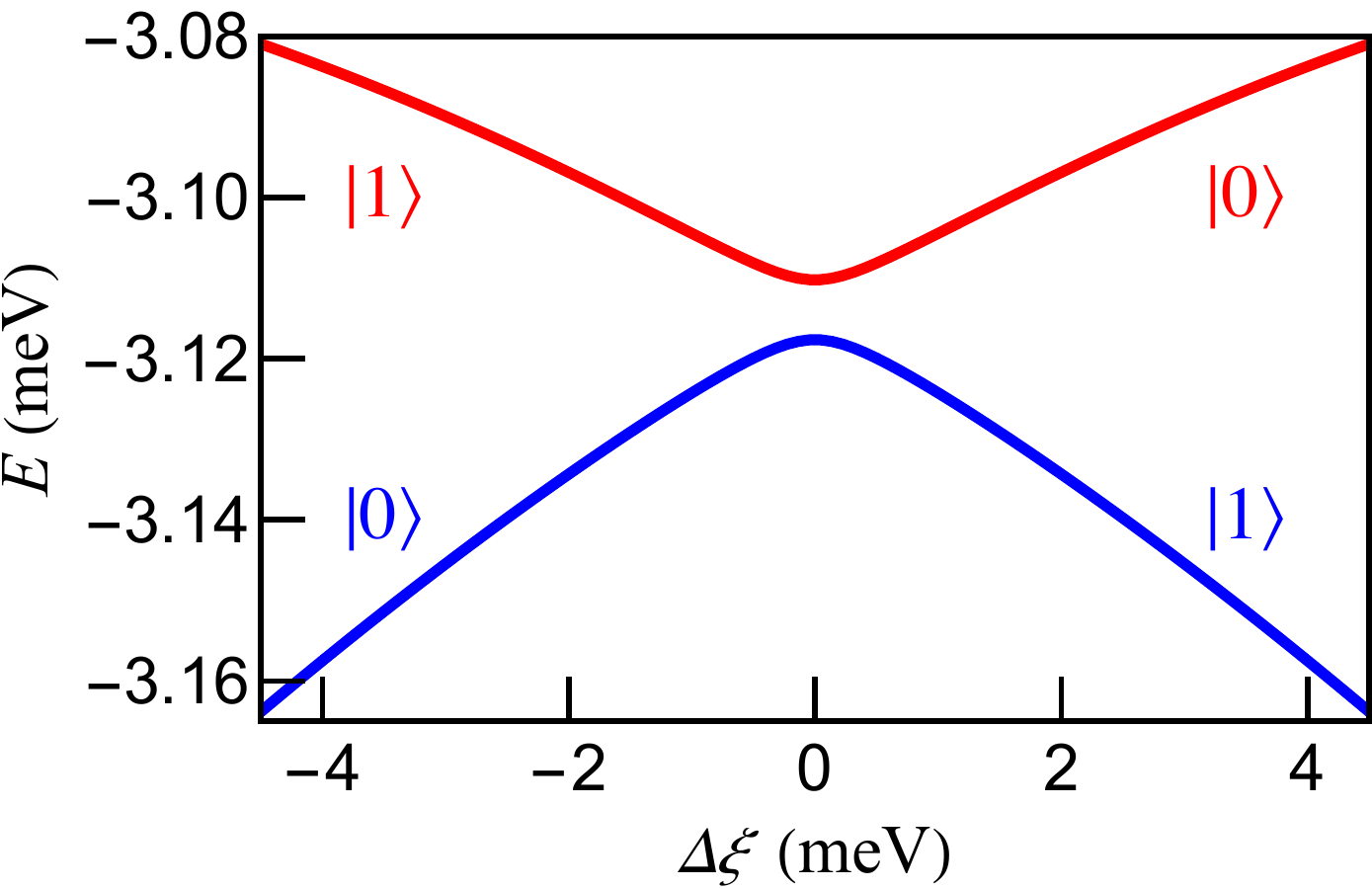}
\hspace{\fill}
(b)\centering\includegraphics[width=0.7\columnwidth]{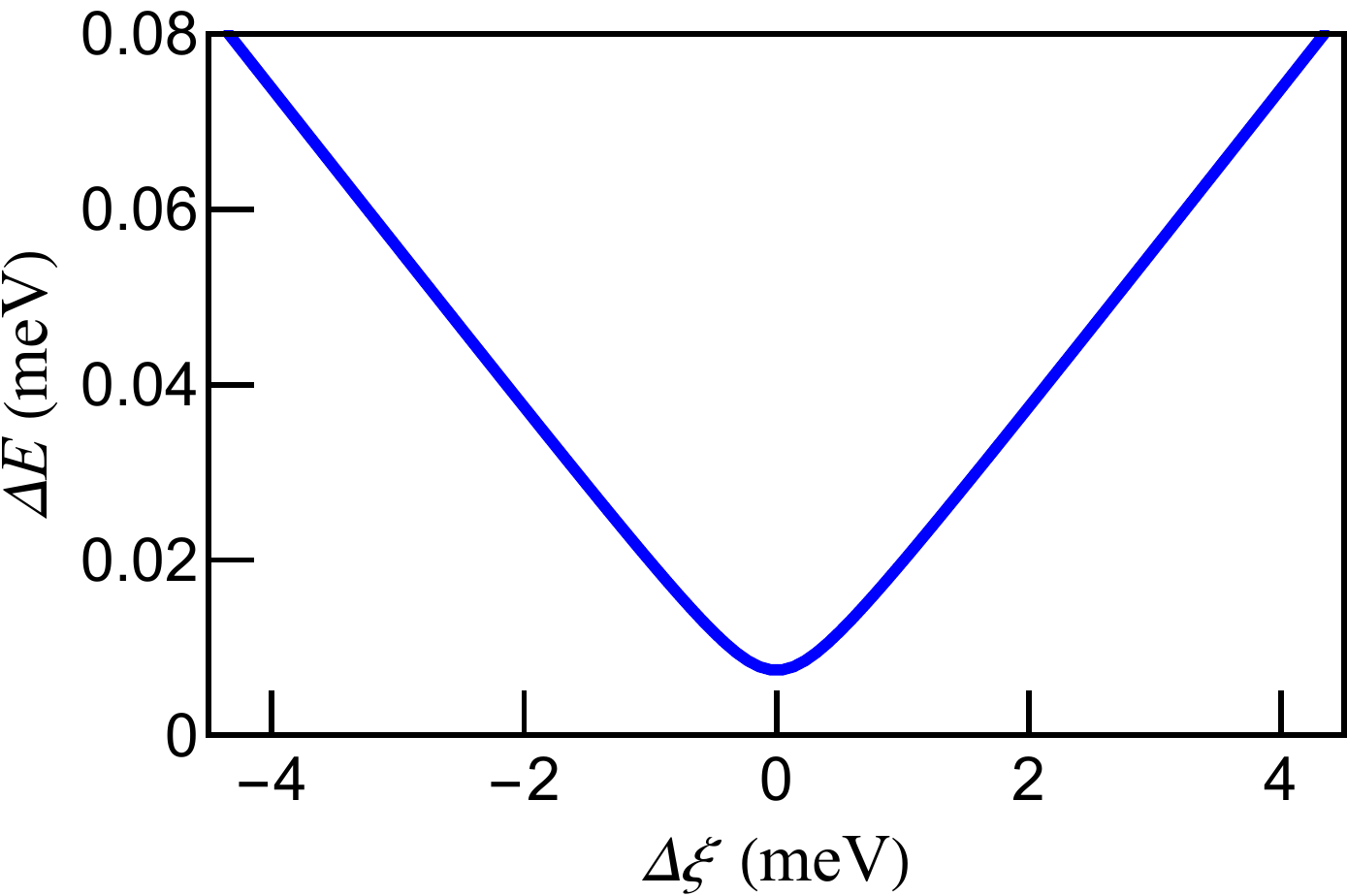}
\caption{(a) Calculated energy spectra of the triple-dot charge qubit system as functions of $\Delta\xi=\xi_1-\xi_2$. Only the lowest two energy levels are shown.  (b) The energy level splitting between the two states shown in (a) as a function of $\Delta\xi$. Parameters: $\xi_1+\xi_2=4.5$ meV.  $a$ = 150 nm,  $\hbar\omega_0$ = 120 $\mu$eV, $\mu_1$ = $\mu_3$ = 0 and $\mu_2$ = -4 meV.}
\label{fig:spectradxi}
\end{figure}

\begin{figure}
(a)\centering\includegraphics[width=0.7\columnwidth]{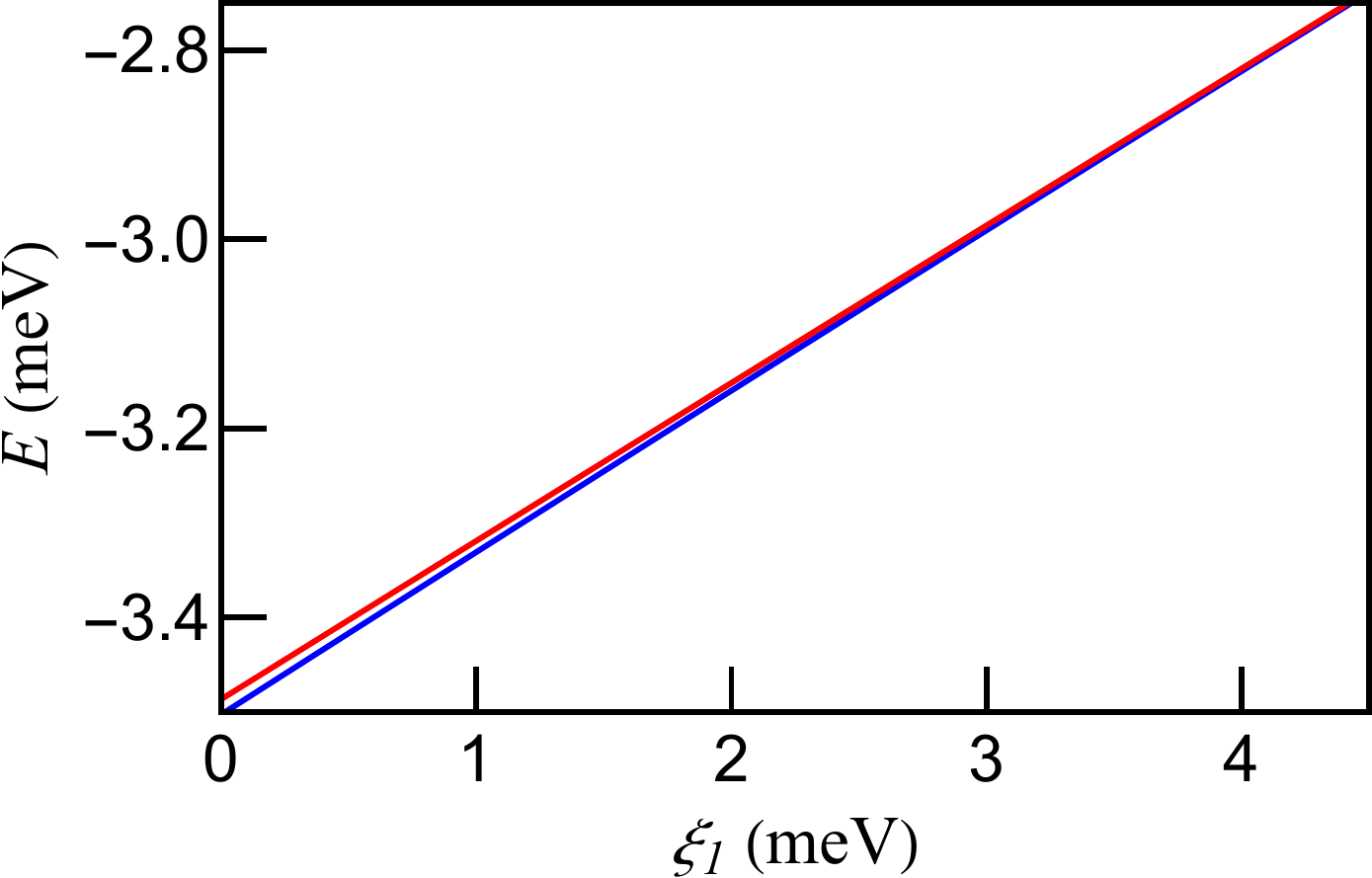}
\hspace{\fill}
(b)\centering\includegraphics[width=0.7\columnwidth]{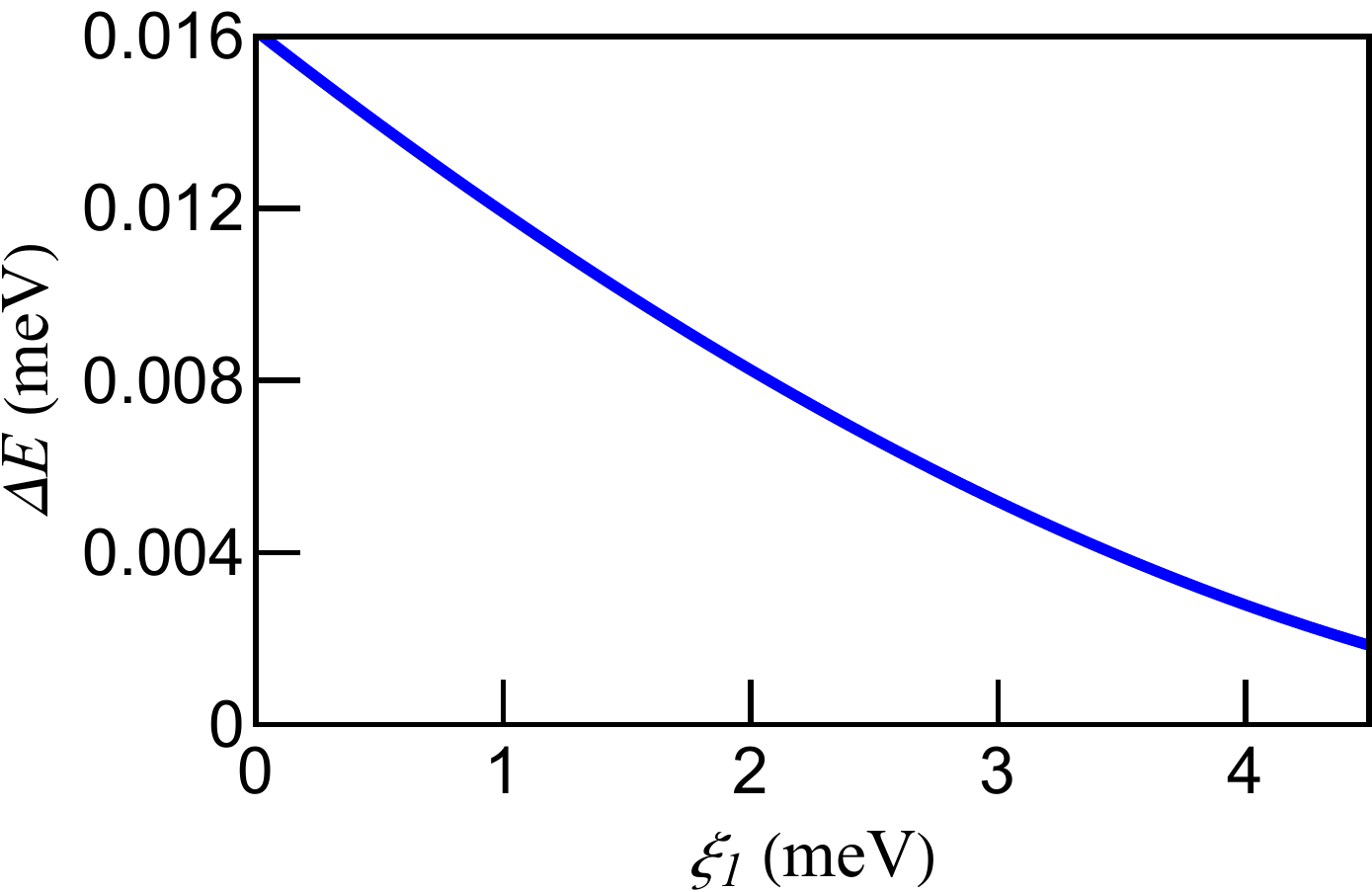}
  \caption{(a) Calculated energy spectra of the triple-dot charge qubit system as functions of $\xi_1$, where $\xi_1=\xi_2$. Only the lowest two energy levels are shown.  (b) The energy level splitting between the two states shown in (a) as a function of $\xi_1$. Parameters: $a$ = 150 nm,  $\hbar\omega_0$ = 120 $\mu$eV, $\mu_1$ = $\mu_3$ = 0 and $\mu_2$ = -4 meV.}
\label{fig:spectraxi1}
\end{figure}

Figure~\ref{fig:spectradxi}(a) shows the calculated energy spectra of our system as functions of the difference in the barrier heights, $\Delta\xi=\xi_1-\xi_2$, while $\xi_1+\xi_2$ is fixed at 4.5 meV. Only the lowest two energy levels are shown as others are well separated from the qubit subspace. The form of the curves is typical for a charge qubit, namely the energies of the two qubit states avoid crossing in the middle. The direction of the rotation axis is varied as $\Delta\xi$ is changed from negative to positive values, as will be demonstrated below. The difference between the two energy levels, $\Delta E$ gives the rotation speed around the concerned axis. For $\xi_1+\xi_2=4.5$ meV, $\Delta E$ as a function of $\Delta\xi$ is shown in  Fig.~\ref{fig:spectradxi}(b). Changing $\xi_1+\xi_2$ varies the amplitude of $\Delta E$, thus tunes the rotating speed as desired, as shall be demonstrated below.

Figure~\ref{fig:spectraxi1}(a) shows the energies of the qubit states as functions of $\xi_1$, where keeping $\xi_1=\xi_2$. Both energies are changed as the barrier heights are varied (in this case, the two barriers have to be varied simultaneously). The difference between them, $\Delta E$ as a function of $\xi_1$, is shown in Fig.~\ref{fig:spectraxi1}(b). When $\xi_1$ and $\xi_2$ are raised from 0 to 4.5 meV, $\Delta E$ is reduced from around 0.016 meV to 0.002 meV, namely by a factor of eight. $\Delta E$ is smallest at $\Delta\xi=0$, and we have found that it is even better tunable at more positive and negative $\Delta\xi$ values.

\begin{figure}
(a)\centering\includegraphics[width=0.7\columnwidth]{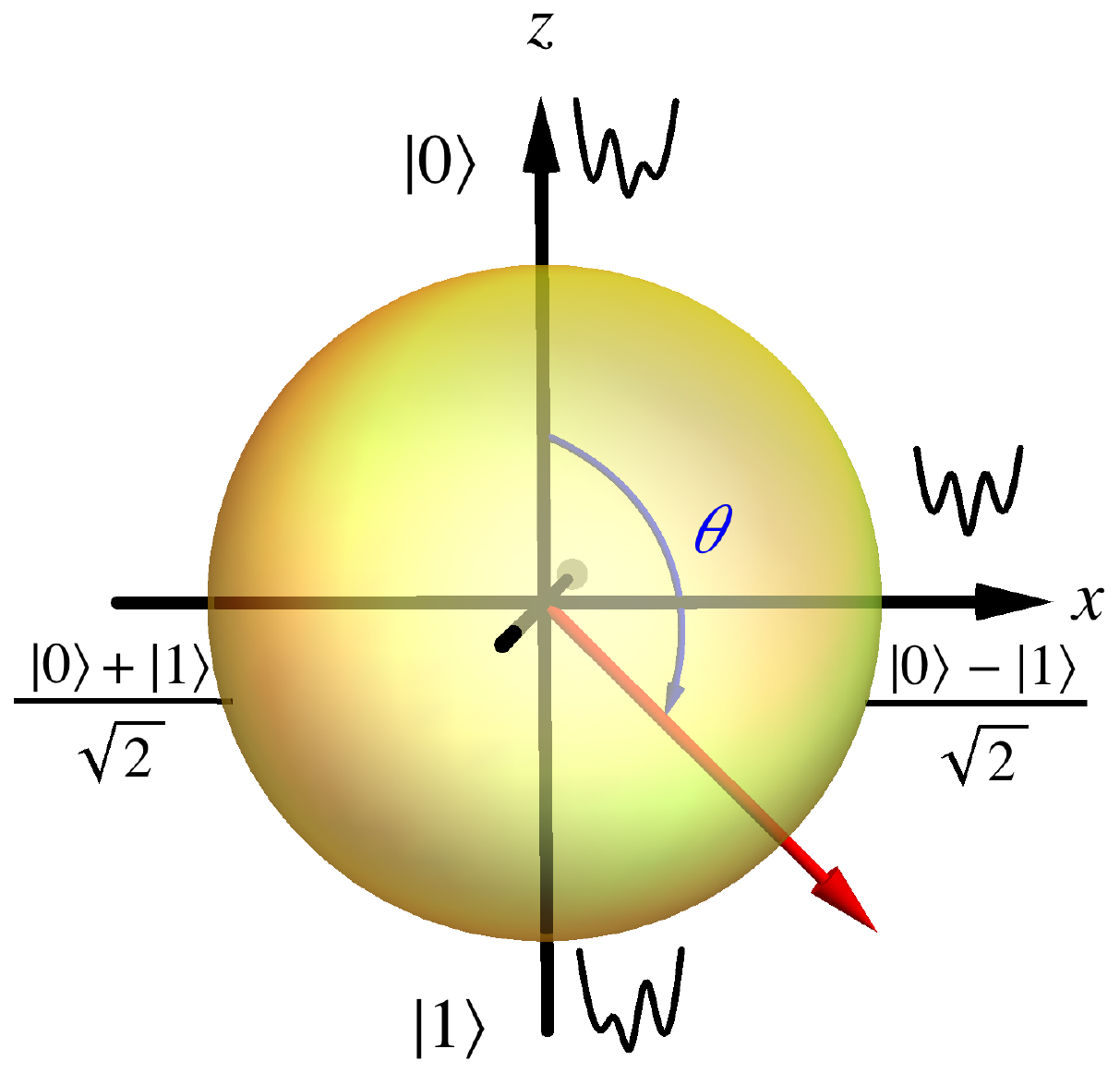}
\hspace{\fill}
(b)\centering\includegraphics[width=0.7\columnwidth]{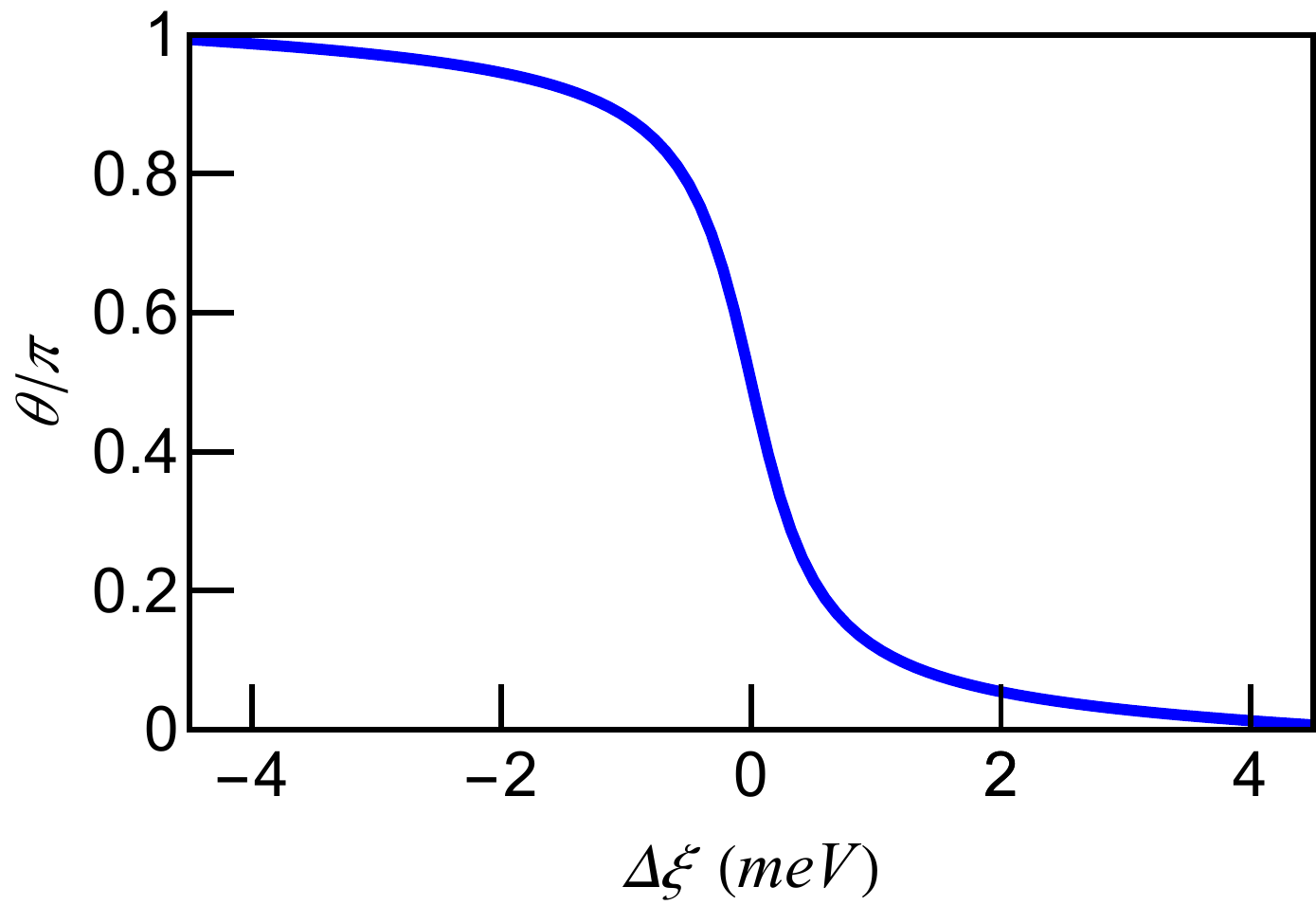}
  \caption{(a) The direction of the rotation axis controlled by the relative height of two barriers. The rotation axis is represented by a red arrow that is in general an angle $\theta$ apart from $\hat{z}$. It is also noted that when $|\Delta\xi|$ is large, the rotation axis is close to $\pm\hat{z}$, while when $\Delta\xi=0$, the axis is along $\hat{x}$. (b) The angle between the rotation axis and $\hat{z}$ as a function of $\Delta\xi$. Parameters: $\xi_1+\xi_2=4.5$ meV, $a = 150$nm, $\hbar\omega_0$ = 120 $\mu$eV, $\mu_1$ = $\mu_3$ = 0 and $\mu_2 = -4$ meV.}
\label{fig:direction}
\end{figure}

Figure~\ref{fig:direction} shows how the rotation axis is varied with $\Delta\xi$. The axis is found by projecting the eigenstates $(\left|\psi\right)$ corresponding to a specific $\Delta\xi$ value to the qubit bases, i.e.
\begin{equation}
\left|\psi\right \rangle=\cos\frac{\theta}{2}\left|0\right \rangle+\sin\frac{\theta}{2}\left|1\right \rangle.
\label{eq:q}
\end{equation}
Therefore the rotation axis is in the $xz$ plane apart from $\hat{z}$ by an angle \cite{leakagenote}
\begin{equation}
\theta=2\arctan\frac{\langle1|\psi\rangle}{\langle0|\psi\rangle}.
\label{eq:theta}
\end{equation}
Fig.~\ref{fig:direction}(a) schematically depicts how the relative height of the two barriers affects the direction of the  rotation axis. When the two barriers are vastly different in height, the rotation axis is close to $\hat{z}$; on the other hand, when the two barriers are leveled, the rotation axis is $\hat{x}$. The detailed dependence of $\theta$ on $\Delta\xi$ is shown in  Fig.~\ref{fig:direction}(b). When $\Delta\xi<0$, $\theta$ decreases from $\pi$ and approaches $\pi/2$. $\theta=\pi/2$ for $\Delta\xi=0$. $\theta$ decreases to around $0$ when $\Delta\xi$ further increases.

\begin{figure}
  \includegraphics[width=0.75\columnwidth]{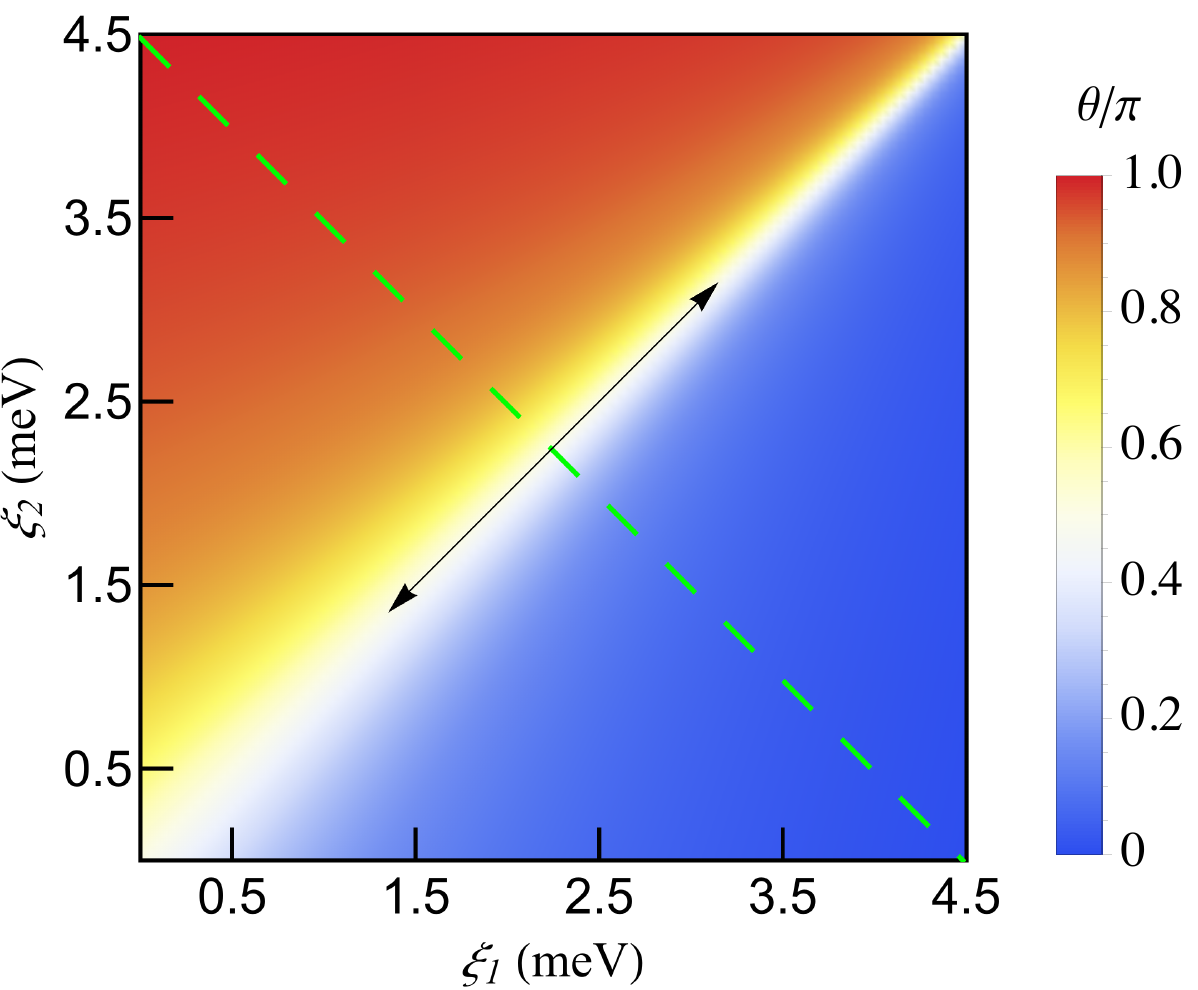}
  \caption{Pseudocolor plot of the angle between the rotation axis and $\hat{z}$, $\theta$, as a function of $\xi_1$ and $\xi_2$. $\xi_1+\xi_2=4.5$ meV has been marked as the green/gray dashed line, while the double arrow indicates directions along which the amplitude of the qubit energy level splitting can be tuned.  Parameters are: $a$ = 150 nm,  $\hbar\omega_0$ = 120 $\mu$eV, $\mu_1$ = $\mu_3$ = 0 and $\mu_2 = -4$ meV.}
\label{fig:psucolor}
\end{figure}

Fig.~\ref{fig:psucolor} is a pseudocolor plot of the angle between the rotation axis and $\hat{z}$, $\theta$, as a function of both $\xi_1$ and $\xi_2$. In this figure, $\xi_1+\xi_2=4.5$ meV has been marked as the green/gray dashed line, corresponding to the parameter range used in Figs.~\ref{fig:spectradxi}  and \ref{fig:direction}. Along this direction, the rotation axis can be varied. The double arrow along $\xi_1=\xi_2$ indicates the directions along which the amplitude of the qubit energy level splitting can be changed, offering additional tunability to the qubit system.

\begin{figure}[t]
(a)\centering\includegraphics[width=0.4\columnwidth]{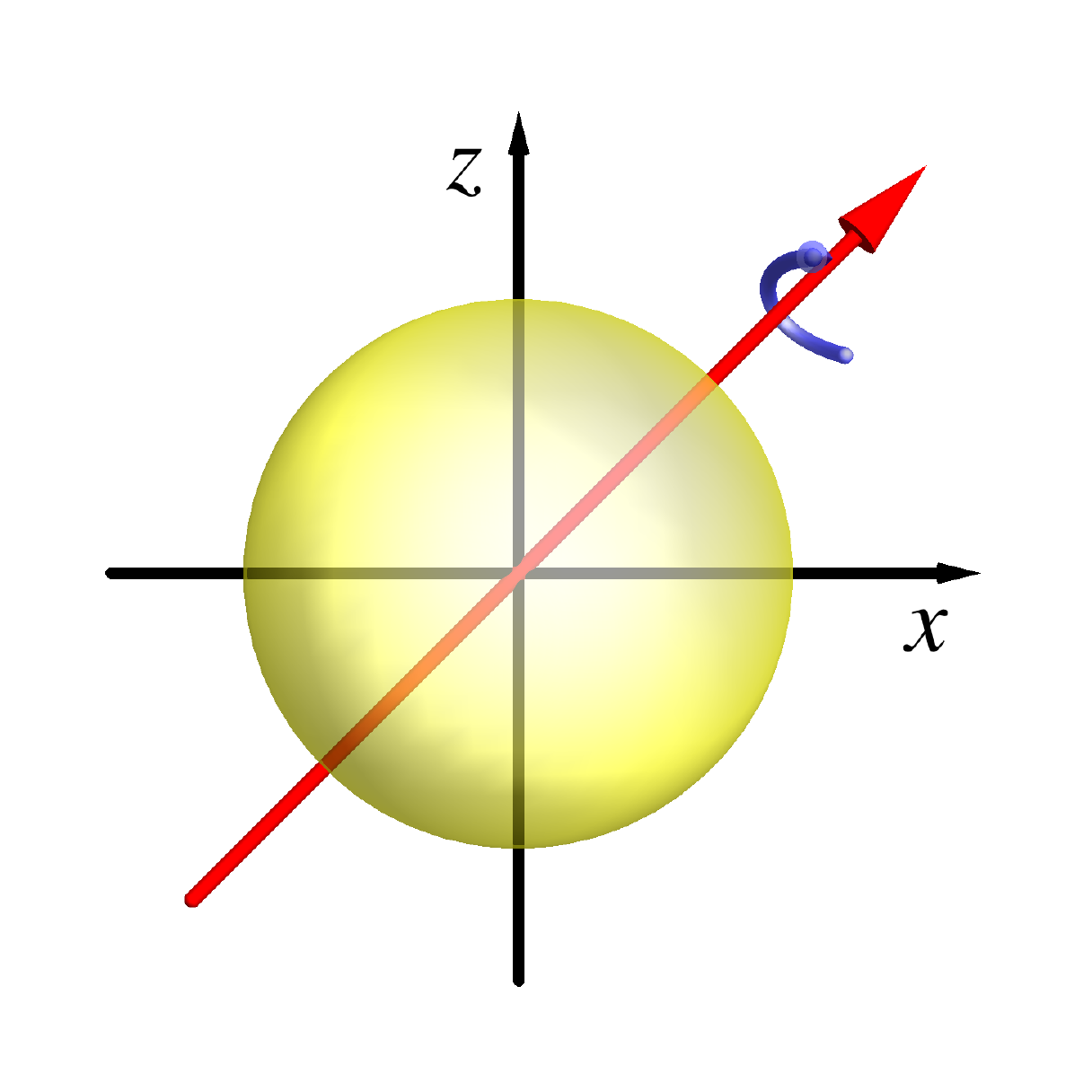}
(b)\centering\includegraphics[width=0.4\columnwidth]{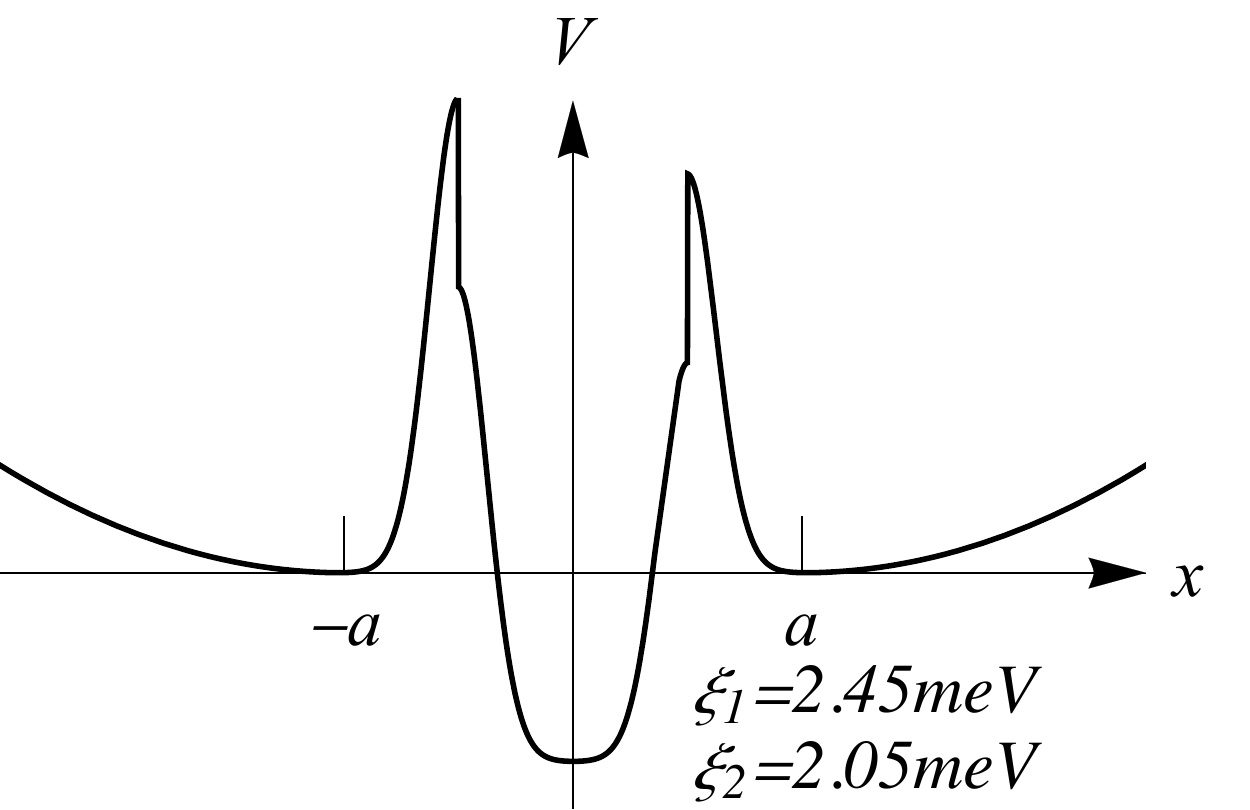}
\hspace{\fill}
(c)\centering\includegraphics[width=0.4\columnwidth]{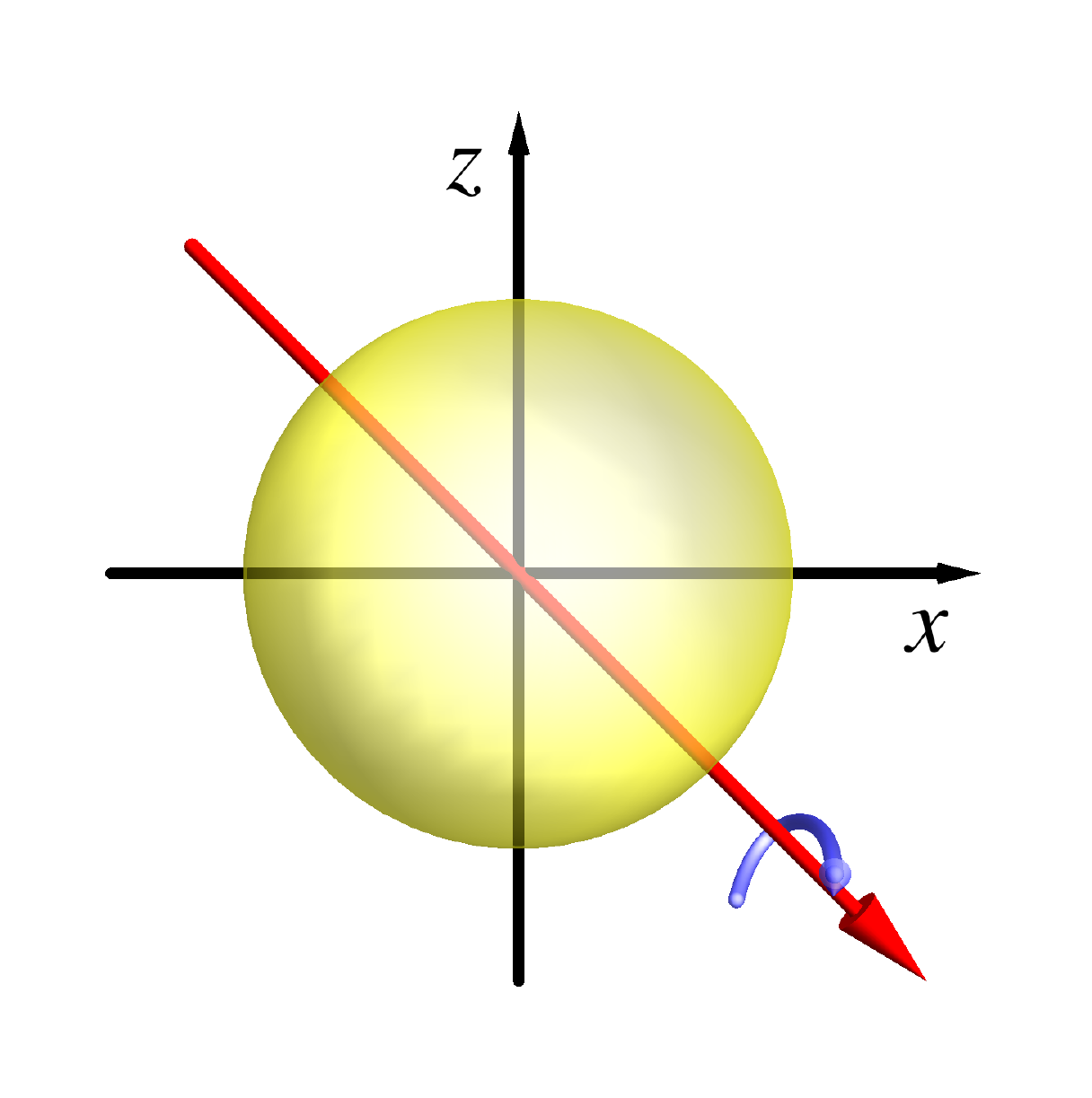}
(d)\centering\includegraphics[width=0.4\columnwidth]{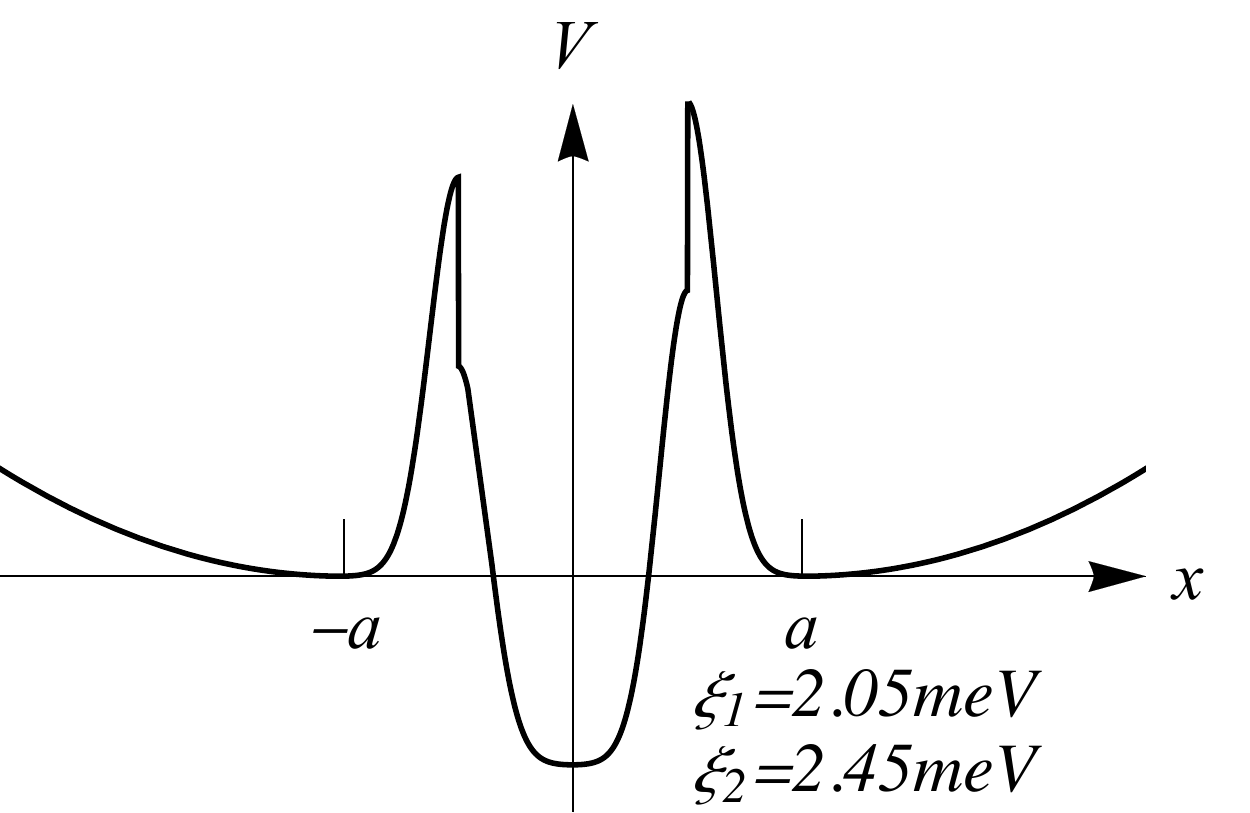}
\caption{Scenarios with $\theta=\pi/4, 3\pi/4$. (a) Bloch sphere showing rotation axis at $\theta=\pi/4$. (b) Potential profile for (a) with $\xi_1$ and $\xi_2$ indicated.  (c) Bloch sphere showing rotation axis at $\theta=3\pi/4$. (b) Potential profile for (c) with $\xi_1$ and $\xi_2$ indicated. The parameters are: $a = 200$ nm,  $\hbar\omega_0 = 60$ $\mu$eV, $\mu_1 = \mu_3 = 0$ and $\mu_2 = -4$ meV.}
\label{fig:scenarios}
\end{figure}

As two examples of the qubit operation, we show the rotation axes $\hat{x}+\hat{z}$ and $\hat{x}-\hat{z}$ $(\theta=\pi/4,3\pi/4)$ and the corresponding confinement potential in Fig.~\ref{fig:scenarios}. Fig.~\ref{fig:scenarios}(a) and (b) show the case for rotation axis $\hat{x}+\hat{z}$. Fig.~\ref{fig:scenarios}(a) is a schematic plot showing the axis on the Bloch sphere, while Fig.~\ref{fig:scenarios}(b) shows the shape of the confinement potential on the $xz$ plane along with the parameters $\xi_1=2.45$ meV and $\xi_2=2.05$ meV used. Note that the barrier at $x=-a$ is higher than the one at $x=a$. Fig.~\ref{fig:scenarios}(c) depicts the rotation axis $\hat{x}-\hat{z}$ on the Bloch sphere, while Fig.~\ref{fig:scenarios}(d) shows the confinement potential on $xz$ plane with  $\xi_1=2.05$ meV and $\xi_2=2.45$ meV. In this case, the barrier at $x=a$ is higher than the one at $x=-a$.

\blue{To show the feasibility of our proposed qubit encoding scheme in triple quantum dots, we proceed with numerical simulations on the time evolution of the qubit state for two representative single-qubit gates, the Rabi oscillation (x-rotation) and the Hadamard gate (rotation around $\hat{x} + \hat{z}$). 
}
\begin{figure} [tbp]
	\centering{
	\includegraphics[width=0.8\columnwidth]{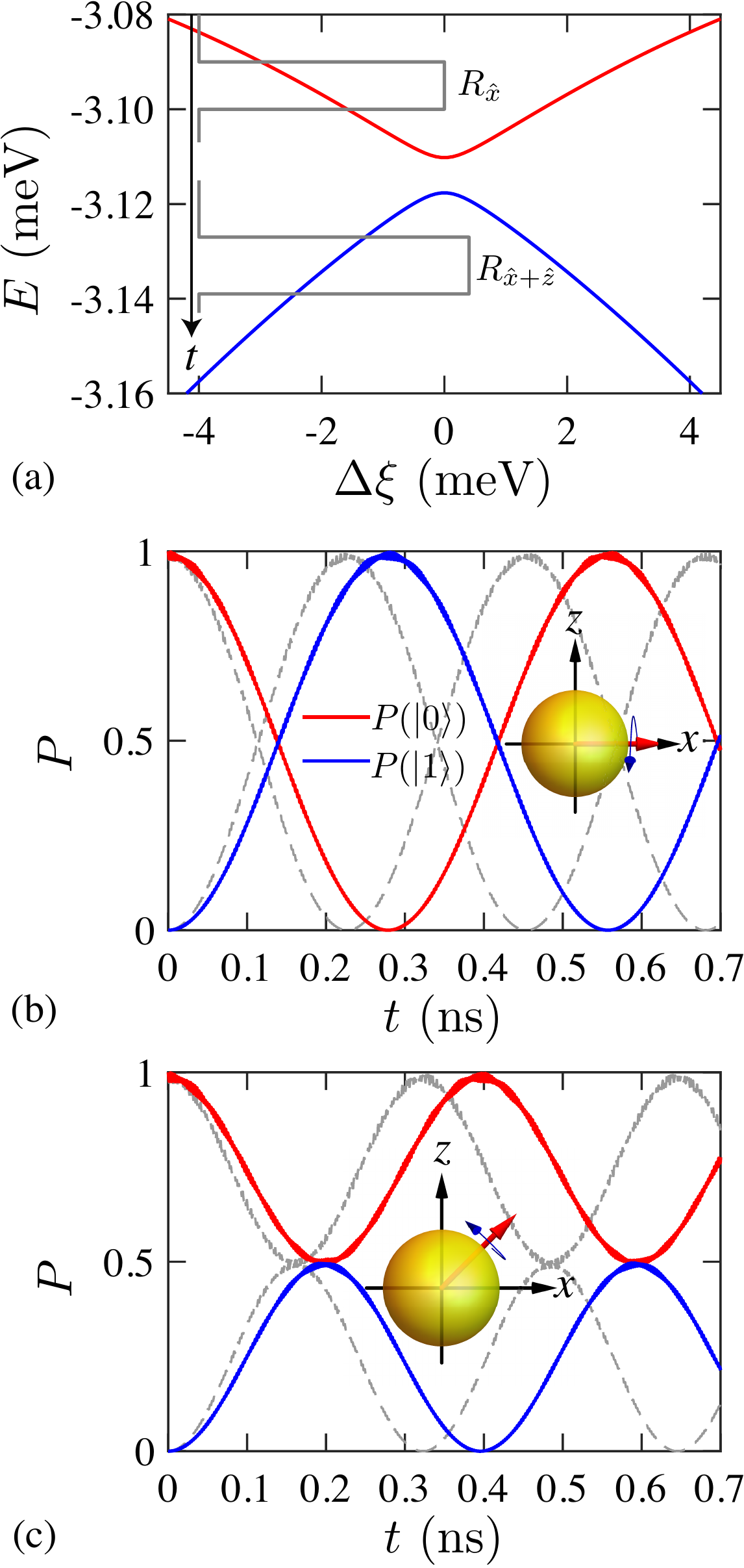}
	}
	\caption{\blue{(a) The same calculated energy spectra of the triple-dot charge qubit system as Fig.~\ref{fig:spectradxi}(a), superposed with schematic pulse shapes for performing Rabi oscillation and Hadamard gates. (b) Numerical simulations for Rabi oscillation ($\Delta \xi \approx 0$ meV). (c) Numerical simulations for the Hadamard gate ($\Delta \xi = 0.401$ meV)). Red (Blue) lines correspond to the probability of $|0\rangle$ ($|1\rangle$) as functions of time where $\xi_1 + \xi_2 = 4.5$ meV. The corresponding gray dashed lines are the state probabilities when $\xi_1 + \xi_2 = 3.5$ meV ($\Delta \xi \approx 0$ for Rabi Oscillation while $\Delta \xi = 0.504$ meV). Parameters: $a = 150$ nm,  $\hbar\omega_0 = 120$ $\mu$eV, $\mu_1 = \mu_3 = 0$ and $\mu_2 = -4$ meV.}}
	\label{fig:numericSim}
\end{figure}

\blue{Figure~\ref{fig:numericSim} shows the probabilities of states $|0\rangle$ and $|1\rangle$ as functions of time, for Rabi oscillation and the Hadamard gate respectively. The single-qubit gates can be operated by pulsing the relative barrier height, $\Delta \xi$, from a maximum, where $|0\rangle$ is the ground state, to the desired value. When $\Delta \xi = 0$, in the case where $\xi_1 + \xi_2 =4.5$ meV, we can perform the Rabi oscillation between $|0\rangle$ and $|1\rangle$ with frequency $f = 1.795$ GHz (period $T=0.557$ ns) (cf. red and blue lines in Fig.~\ref{fig:numericSim}(b)). Also, when $\Delta \xi = 0.401$ meV ($\xi_1 + \xi_2 =4.5$ meV), the Hadamard gate operation is performed on the qubit state, in which the initial state $|0\rangle$ is unitarily transformed into $(|0\rangle + |1\rangle)/\sqrt{2}$ with $f=2.528$ GHz ($T=0.396$ ns) (cf. red and blue lines in Fig.~ \ref{fig:numericSim}(c)). The technique for pulsing the barrier height nonadiabatically has been achieved in experiments where the tuning of barrier heights can be observed to range from as low as $10$ mV \cite{Malinowski.17} to $110$ mV $\sim190$mV \cite{Reed.16,Martins.16}.
To emphasize on the simultaneous tunability of both rotation angle and rotation speed, which is the main focus of our paper, the above mentioned gate operation on the qubit state is again numerically simulated with a smaller $\xi_1+\xi_2$, (i.e. $\xi_1+\xi_2=3.5$ meV). Evidently, both Rabi oscillation and Hadamard gate operation are observed (cf. the gray lines, in correspondence with red and blue lines, in Fig.~\ref{fig:numericSim} (b) and (c)), of which the frequency $f=2.203$ GHz ($T=0.454$ ns) for the former while $f=3.098$ GHz ($T=0.323$ ns) for the latter. The gate operations can be performed when $\Delta \xi $ is tuned to $0$ meV and $0.504$ meV respectively while maintaining $\xi_1 + \xi_2 = 3.5$ meV. Our simulation results are consistent with the typical oscillation frequency observed in charge qubit schemes where Rabi oscillation frequency ranges from $\sim1$ GHz to $\sim5$ GHz \cite{Scarlino.17,Wang.17CQ,Kim.15,Shi.14,Shi.13}. In addition, our proposed qubit scheme allows additional controllability of the oscillation frequency which is absent in the conventional charge qubit. Before we end this section, we would like to bring the readers' attention to the small-amplitude, high-frequency oscillation of the curves shown in Fig.~\ref{fig:numericSim}(b) and (c): these are due to the leakage out of the qubit subspace, which is a rather small effect. This leakage effect will be further discussed in Appendix~\ref{appx:threelevels}.
}

\section{Conclusions}
\label{sec:conclusion}

In this paper, we have demonstrated a tunable charge qubit based on triple quantum dots. While the energy of all three dots are fixed, the manipulation is performed using the two barriers between the three dots. When the relative height of the two barriers are changed, the rotation axis for single-qubit operation is varied so as to offer flexibility in performing quantum algorithms. Moreover, when both barriers are raised or lowered together, the amplitude of the qubit energy level splitting is altered, and so as the rotating speed. \blue{We have performed numerical simulations and have shown that the rotation angle and gate duration can be individually tuned. Our proposal eliminates the need of microwave pulses to perform rotations, which potentially possesses the risk of heating the sample and causing dephasing, by simply replacing it with non-adiabatic barrier height pulses. Our proposal allows for tunability of both the rotation axis and rotating speed via all electrical control, which may be an alternative method to realize quantum algorithms in these devices.}

This work is supported by the
Research Grants Council of the Hong Kong Special Administrative Region, China (No.~CityU 21300116, CityU 11303617), the National Natural Science Foundation of China (No.~11604277), and the Guangdong Innovative and Entrepreneurial Research Team Program (No.~2016ZT06D348).

%\onecolumngrid

\appendix
\setcounter{equation}{0}

\section{\blue{Results beyond the Hund-Mulliken approximation}}
\label{appx:threelevels}

\blue{The results presented in the main text is discussed based on the Hund-Mulliken approximation, where only the ground state of each quantum dot is included in the calculation. Here, we would like to investigate the applicability of this approximation.

\begin{figure} [t]
	\includegraphics[width=0.9\columnwidth]{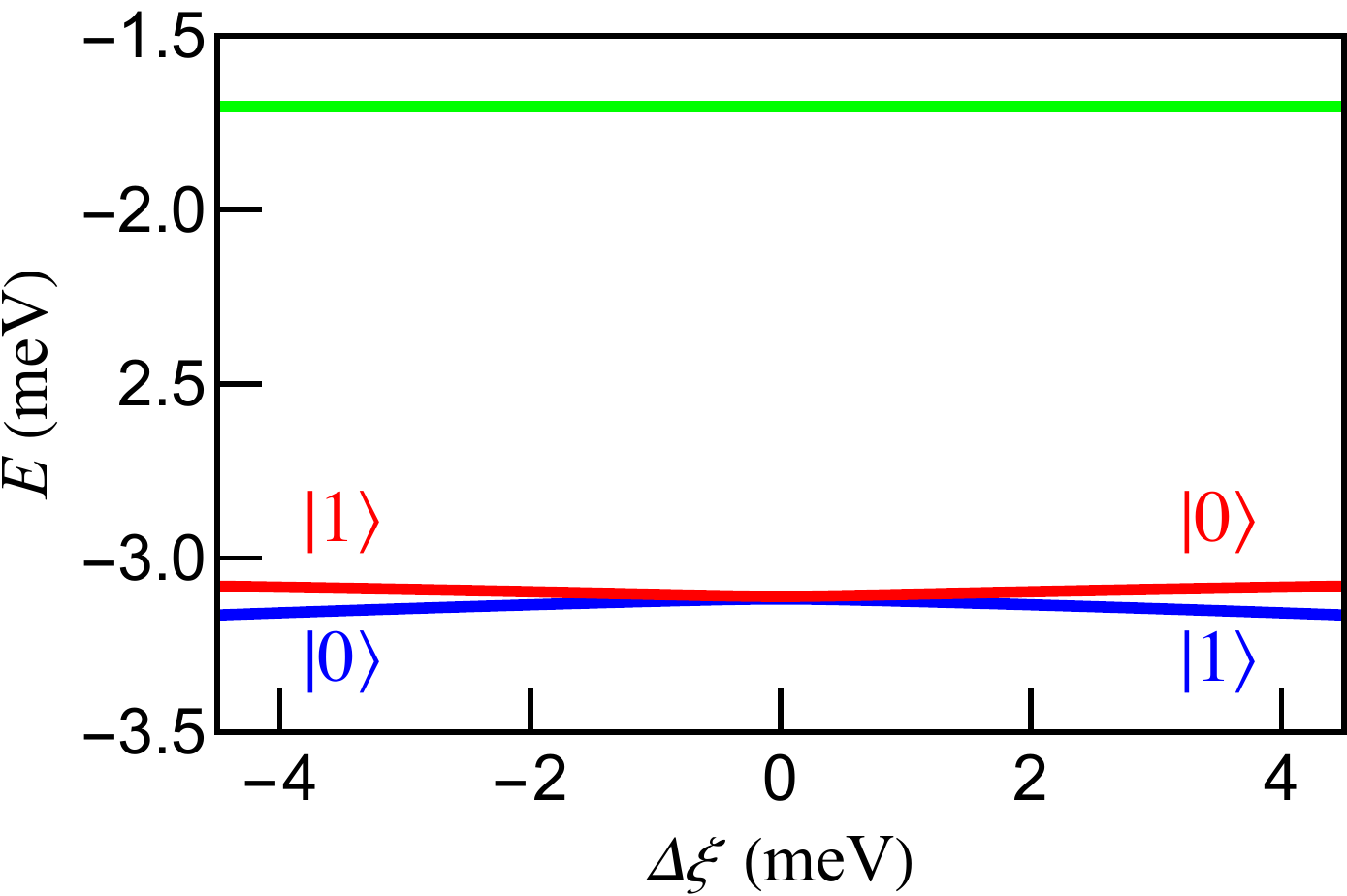}
	\caption{Calculated energy spectra (with three lowest energy levels shown) of the triple-dot charge qubit system as functions of $\Delta \xi = \xi_1 - \xi_2$ under the Hund-Mulliken approximation.  The results are identical to that shown in Fig~\ref{fig:spectradxi}(a). Parameters: $\xi_1+\xi_2=4.5$ meV.  $a$ = 150 nm,  $\hbar\omega_0$ = 120 $\mu$eV, $\mu_1$ = $\mu_3$ = 0 and $\mu_2=-$4 meV.}
	\label{fig:appenHMThreeEval}
\end{figure}

Fig.~\ref{fig:appenHMThreeEval} shows the Hund-Mulliken results with lowest three energy levels of the triple-dot change qubit, among which the lowest two eigenstates are encoded as our logical bases. The fact that the second excited eigenstate is far away in energy from the computational bases means that under this approximation, the quantum computation is hardly being affected by higher-lying orbitals. This fact is also consistent with the calculated leakage $(<0.8\%)$ out of qubit subspace, which can be neglected.

\begin{figure} [htbp]
	\centering{
	\includegraphics[width=1\columnwidth]{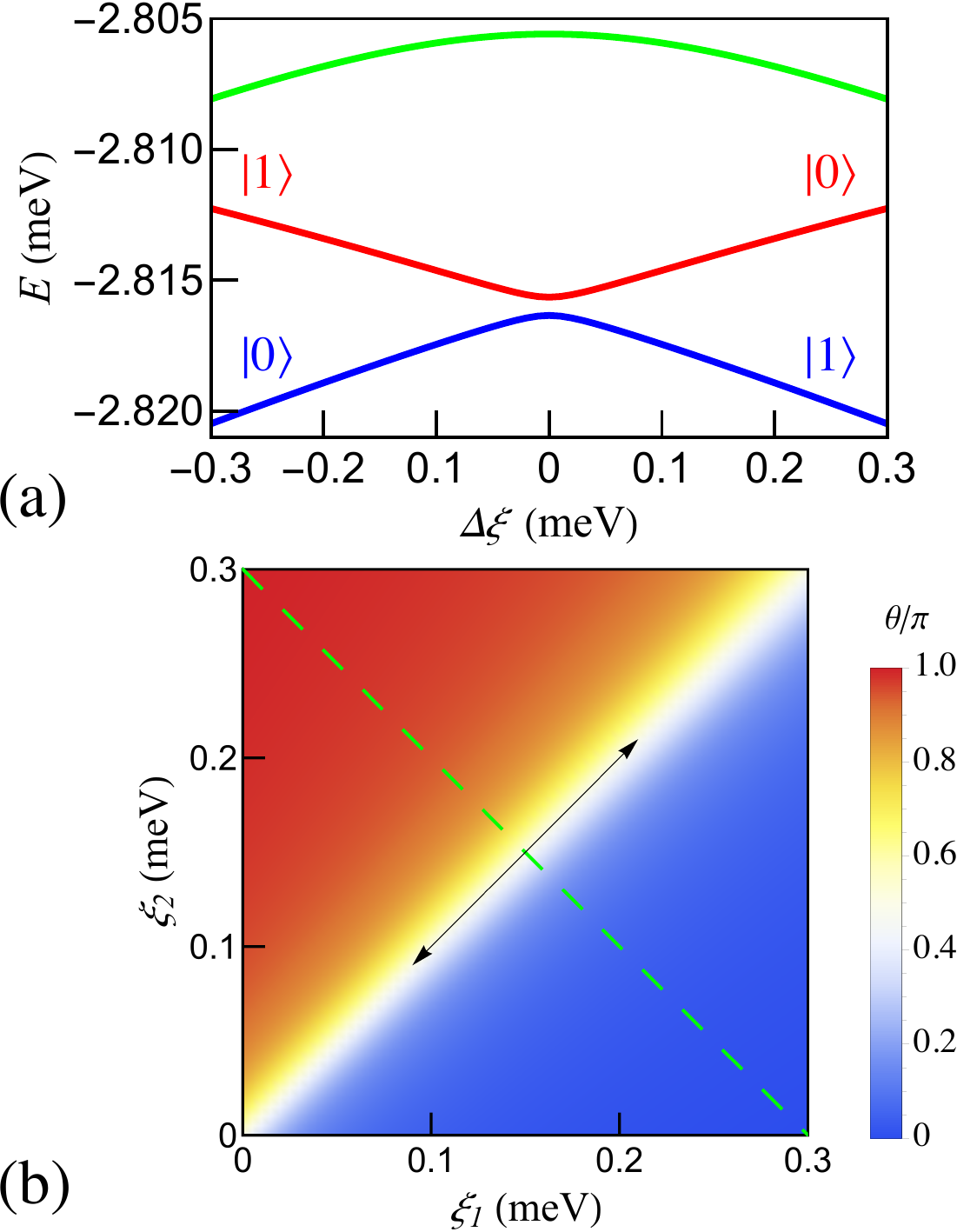}
	}
	\caption{(a) Calculated energy spectra of the triple-dot charge qubit as functions of $\Delta \xi = \xi_1 - \xi_2$ in which three lowest levels in each quantum-dot are included in the calculation. (b) Pseudocolor plot of the angle between the rotation axis and $\hat{z}$, $\theta$, as a function of $\xi_1$ and $\xi_2$. $\xi_1+\xi_2=0.3$ meV has been marked as the green/gray dashed lines, while the double arrow indicates directions along which the amplitude of the qubit energy level splitting can be tuned. Parameters: $\xi_1+\xi_2=0.3$ meV.  $a$ = 210 nm,  $\hbar\omega_0$ = 220 $\mu$eV, $\mu_1$ = $\mu_3$ = 0 and $\mu_2=-$2.3 meV.}
	\label{fig:appenThreeLevelPerDotThreeEval}
\end{figure}

Beyond the Hund-Mulliken approximation, one has to keep at least three orbitals per dot. Next to the ground state (usually called the $s$ orbital, with principal quantum number $n=0$ and magnetic quantum number $m=0$), there are two  $p$ orbitals ($n=1$, $m=\pm 1$). We have performed calculations keeping these three orbitals per dot with the similar dot parameters as has been used in the previous Hund-Mulliken approximation ($\hbar \omega_0=220$ $\mu$eV, $a=210$ nm). Fig.~\ref{fig:appenThreeLevelPerDotThreeEval}(a) shows the calculated lowest three levels when considering three lowest orbitals  in each dot and we are able to recover the same qualitative result as compared to the results under the Hund-Mulliken approximation (cf. Fig.~\ref{fig:appenHMThreeEval}). Fig.~\ref{fig:appenThreeLevelPerDotThreeEval}(b) shows the rotation angle as a function of $\xi_1$ and $\xi_2$, similar to Fig.~\ref{fig:psucolor}. Again, we observed comparable qualitative results for which both rotation axis and rotating speed can be individually tuned by selecting specific $\Delta \xi$ (rotation angle) and $\xi_1 + \xi_2$ (rotating speed). The main difference between results from the Hund-Mulliken approximation (Fig.~\ref{fig:psucolor}) and results keeping three orbitals per dot (Fig.~\ref{fig:appenThreeLevelPerDotThreeEval}) is the specifications of the dot parameters, which are merely based on surface gates' voltage manipulation. In practical experiments, one should be able to resolve the gate voltage required in the tuning process as there are no qualitative difference from our proposal from a simplified model. The calculated leakage out of qubit state is $\lesssim1.01\%$ and the energy difference, $\Delta E$, between the lowest two eigenstates and second excited state is $( E_3 - \frac{E_1 + E_2}{2}) / \Delta E_{12} \approx 2$ (where $E_i$ is the eigenvalue of i$^{\text{th}}$ eigenstate); with such results, we are confident that higher energy levels should not affect gate operations in any significant way.}


\begin{thebibliography}{67}
\bibitem [{fix()}]{fixedenergylevels}%
  \BibitemOpen
  \href@noop {} {}\bibinfo {note} {We note that the single particles energies
  undergo changes when barrier heights are tuned. Hence, instead of the
  detuning of each quantum dot ($\mu_i$), the matrix elements in the
  Hamiltonian that represent single-particles energies ($\varepsilon_i$, where
  $i = 1,2,3$, indicating the electron energy occupying each quantum dot) are
  taken to be the result of integration ($\langle \psi_i | h(\mathbf{r}) |
  \psi_i \rangle$)}\BibitemShut {NoStop}%
\end{thebibliography}
\end{document}